# Tamm Cavity in the THz Spectral Range

Simon Messelot[1], Clémentine Symonds[2], Joël Bellessa[2], Jérôme Tignon[1], Sukhdeep Dhillon,[1] Jean-Blaise Brubach[3], Pascale Roy[3], Juliette Mangeney[1,*]

[1]Laboratoire de Physique de l'Ecole normale supérieure, ENS, Université PSL, CNRS, Sorbonne Université, Université de Paris, F-75005 Paris, France

[2] Institut Lumière Matière, Université Claude Bernard Lyon 1, CNRS, , F-69622, Lyon, France

[3]Synchrotron SOLEIL, L'Orme des Merisiers Saint-Aubin, F-91192 Gif sur Yvette, France

*Corresponding author: *juliette.mangeney@ens.fr*

**ABSTRACT:** Electromagnetic resonators, which are based on optical cavities or electronic circuits, are key elements to enhance and control light-matter interaction. In the THz range, current optical cavities exhibit very high-quality factors with $(\lambda/2)^3$ mode volumes limited by diffraction, whereas resonant electronic circuits show low quality factor but provide strong subwavelength effective volume ($10^{-6}\lambda^3$). To overcome the limitations of each type of resonator, great efforts are being devoted to improving the performances of current methods or to the emergence of original approaches. Here, we report on an optical resonator based on Tamm modes newly applied to the THz range, comprising a metallic layer on a distributed Bragg reflector and demonstrating a high-quality factor of 230 at ~1THz. We further experimentally and theoretically show a fine-tuning of the Tamm mode frequency (over a 250 GHz range) and polarization sensitivity by subwavelength structuration of the metallic layer. Electromagnetic simulations also reveal that THz Tamm modes are confined over a $\lambda/2$ length within the distributed Bragg reflector and can be ideally coupled to both bulk materials and 2D materials. These THz Tamm cavities are therefore attractive as basic building blocks of lasers, for the development of advanced THz optoelectronic devices such as sensitive detectors, high-contrast modulators, narrow filters, and polarizers, as well as for THz cavity quantum electrodynamics in nanostructures.

KEYWORDS: Terahertz, Resonator, Tamm cavity, Distributed Bragg Reflector, Subwavelength metal grating, Equivalent circuit model.

## Introduction

Enhancing and controlling the interaction of THz light with various material systems using THz resonators are of paramount importance for fundamental studies of cavity quantum electrodynamics and for the development of advanced THz devices such as sensors, low-power switches/modulators, narrow filters, sensitive detectors and lasers. Therefore, a large interest is being devoted to THz resonators. For instance, strong light-matter coupling between the cyclotron resonances of a 2D electron gas and THz photons have been reported in THz resonators such as metamaterials and Fabry-Perot cavities, opening very attractive perspectives for cavity quantum electrodynamics.[1,2] The enhanced coupling of THz light to intersubband transitions in heterostructures has been demonstrated in disk patch resonators[3] and in subwavelength metal-dielectric microcavities[4] leading to the first generation of cavity polaritons in the THz frequency range. The integration of quantum wells into sub-wavelength three-dimensional resonant circuit[5,6] and into patch antenna cavity array[7] significantly improves performances of THz detectors. A highly sensitive, compact sensing system has been recently demonstrated based on a silicon photonic crystal cavity (Q>10 000) and resonant tunneling diodes.[8] Also, enhancing the interaction of THz light with 2D materials to overcome the weak interaction between the atom-thick layer and normal incident light has been widely investigated. For instance, efficient graphene-based THz modulators have been demonstrated by incorporating a graphene layer into the cavity of a quantum cascade laser[9] or within a Fabry–Perot resonator.[10] Moreover, recent theoretical investigations have predicted enhanced absorption at THz frequencies by a graphene monolayer coated on a one-dimensional photonic crystal, caused by the excitation of optical Tamm states at the interface between the graphene and the photonic crystal.[11] Designs relying on the integration of graphene into a high-quality photonic crystal nanocavity[12] and a high-quality factor distributed feedback structure[13] have also been proposed to realize graphene-based THz lasers. Alternatively, corrugated waveguides with highly reflective photonic crystal mirrors[14] and subwavelength thick Si discs were recently reported[15] as high-quality factor THz resonators for THz gas phase spectroscopy.

All this recent research work uses a wide variety of THz resonators, based either on optical cavities or on electronic circuits, each with their own advantages. For instance, optical cavities, such as Fabry-Perot or photonic crystal cavity exhibit very high-quality factors (>100) with $(\lambda/2)^3$ mode volumes limited by diffraction. In return, electronic circuits, made of miniature metallic structures, show low quality factor (~10) but provide strong subwavelength effective volume ($10^{-6}\lambda^3$). To overcome the limitations of each type of resonator, great efforts are being devoted to improving the performances of current methods or to the emergence of original approaches.[16] Clearly, developing a THz resonator that associates high quality factor and subwavelength confinement still remains an important challenge for the realization of low-loss advanced THz nanodevices and for a strong THz light-matter interaction platform. Here, we report on a new optical cavity in the THz frequency range based on Tamm modes, composed of a distributed Bragg reflector at THz frequencies and a metallic mirror. We demonstrate a high-quality factor of 230 at ~1 THz. Electromagnetic simulations reveal that Tamm modes are confined over a $\lambda/2$ length within the distributed Bragg reflector and can be ideally coupled to both bulk and 2D materials. Furthermore, we experimentally and theoretically show that fine frequency tuning over >250 GHz and polarization sensitivity are achievable in the Tamm cavity with sub-wavelength metal structuration. Our analysis also shows that the polarization dependent frequency tuning is well predicted by an analytical circuit model. Moreover, the Tamm cavity with subwavelength metallic grating acts as a local phase probe, preventing the necessity of complex delay or interferometric measurements.

# THz TAMM CAVITY

## Theoretical Description

Tamm cavities have been widely studied in the past decade in the visible and infrared range.[17,18,19,20,21,22] They are composed of a metal layer covering a distributed Bragg reflector (DBR) made of a stack of high and low refractive index quarter-wavelength layers. The Tamm electromagnetic mode arises at the interface between the metal layer and the periodic stack and is due to the metal inducing a break of periodicity in the structure. In the infrared range, the DBR is typically grown by molecular beam epitaxy or PECVD, such as GaAs/Al$_{0.95}$Ga$_{0.05}$As. These fabrication methods are not directly scalable to the THz range, since the large thickness of quarter-wavelength layers at THz frequencies (~20 μm) is not compatible with epitaxial or PECVD techniques, and THz Tamm cavity schemes have only been proposed theoretically.[23] For the realization of Tamm cavity in the THz spectral range, we use a stack of thin silicon wafers and vacuum layers as an alternation of high and low refractive index quarter-wave layers to form the DBR.[24] This stack ends with the high refractive index layer, a Si layer, covered with a 100 nm thick gold layer (Fig. 1(a)). To center the stop-band of the DBR at 1 THz, 22 μm thick silicon layers are required for making $\lambda_0/4n_{Si}$ layers, with $n_{Si}$ the Si refractive index ($n_{Si}$=3.42 in the THz range) and $\lambda_0$ the vacuum wavelength. However, to overcome any technical issues related to the wafer mechanical fragility, we instead use $3\lambda_0/4n_{Si}$ thick Si layers (so 66 μm thick Si wafers). A DBR made of 66 μm Si and 75 μm vacuum layers shows a well-known photonic stop-band centered at 1 THz, as displayed by the black curve in Fig. 1(b) obtained by Transfer Matrix Method (TMM). This DBR structure acts as an efficient mirror with a bandwidth close to 0.4 THz. By depositing an additional gold mirror on top of the DBR, we build up a zero-thickness optical cavity, the Tamm cavity, which resonates within the DBR stop-band. In Fig. 1(b), we clearly observe a sharp Tamm mode at 1 THz (red curve), with a quality factor as high as 544, a minimum of the reflectivity of 3% and a contrast of the reflectivity resonance of 0.94. The use of 3 times thicker Si wafers compared to usual $\lambda_0/4n_{Si}$ layers has no influence on the cavity resonant frequency. It only induces three small substop-bands instead of one DBR stop-band (0.8 THz wide) resulting in two additional resonant Tamm modes (for details see **Supplementary S1**).

We now turn to the main properties of Tamm modes. The frequency of the Tamm modes is determined by phase matching of a round-trip inside the cavity, similarly to common Fabry-Perot cavities. However, since there is no space between the two mirrors in Tamm cavity, this phase matching condition involves only the complex reflection coefficients of the DBR, $r_{DBR}$, and of the gold mirror, $r_{gold}$.[25] Hence, the phase matching criterion reads:

$$\arg(r_{DBR}) + \arg(r_{gold}) = 2\pi n \quad (1)$$

Gold being an excellent conductor in the THz range, the reflection coefficient phase of the gold mirror is very close to π. Since $\arg(r_{DBR}) = \pi$ at the center of the DBR stop band when the final layer (closest to the gold mirror) is of high refractive index, the resonant frequency of the Tamm cavity falls in the DBR center frequency, as observed in Fig. 1(b).[26] The fundamental frequency of the Tamm mode is then easily predicted and designed as it matches the DBR center frequency. This is in contrast with Tamm cavity in the visible and NIR range, for which resonant frequency is close to stop-band edges of the DBR as the reflection phase on metallic mirrors for these wavelengths is ~0.8π. Note that the very slight deviation (0.1%) from the DBR center frequency seen in Fig. 1(b) is due to the nonperfect reflection of the gold mirror ($r_{gold}$=0.988 at 1 THz). In return, if the final layer is of low refractive index, $\arg(r_{DBR}) = 0$ at the center of the DBR stop band and the frequencies at which $\arg(r_{DBR}) = \pi$ lie outside the DBR stop band, preventing the existence of the Tamm mode.

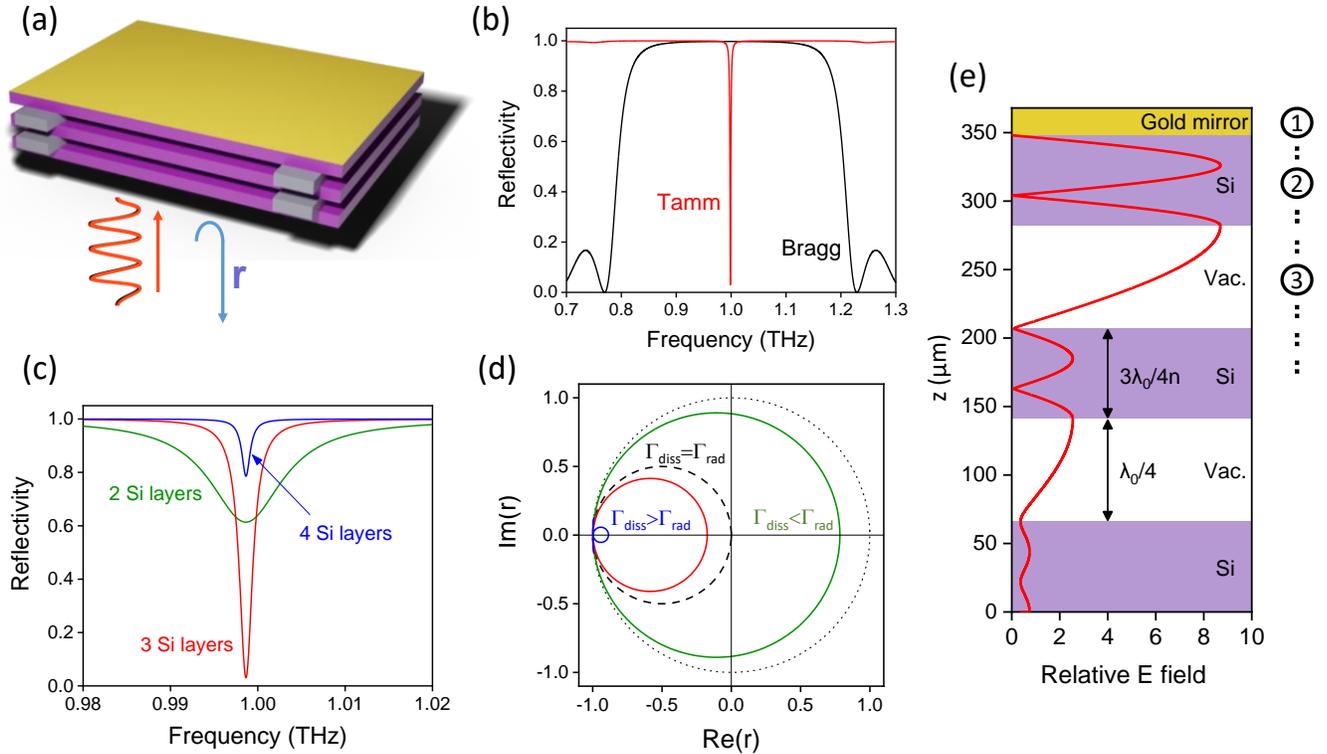

**Fig. 1.** (a) Schematic of the THz Tamm cavity, including a Si-vacuum THz DBR and a gold mirror. (b) Simulated (Transfer Matrix Method, TMM) reflection spectrum of a 3 Si layer-2 vacuum layers DBR (black curve), alongside the corresponding Tamm cavity obtained by adding a gold mirror (red curve). (c) Reflectivity spectra around the resonance peak of the 1 THz Tamm cavity for an increasing number of Si layers in the DBR. The quality factors are 99, 544 and 876 for 2, 3 and 4 silicon layers, respectively. (d) Evolution versus frequency (from 0.96 to 1.04 THz) of the complex reflection coefficient of the Tamm cavity in the complex plan for an increasing number of Si layers. Black dashed line illustrates the hypothetical critical coupling case and gray dotted line represents the unity circle. (e) Structure of the Tamm cavity resonating at 1 THz superimposed to the relative electric field distribution at resonance for normal incidence, computed from TMM.

The quality factor $Q$ of the Tamm cavity is determined by the quality of its mirrors:

$$\frac{1}{Q} = \frac{1}{Q_{DBR}} + \frac{1}{Q_{gold}} \quad (2)$$

We use in the following the radiative coupling rate $\Gamma_{rad}$, the dissipation rate $\Gamma_{diss}$ and superscripts $^1$ and $^2$ for the DBR and the gold mirror, respectively. Due to the use of high-resistivity silicon and vacuum, losses in the DBR are neglectable: $\Gamma_{diss}^1 = 0$ and the contribution of the DBR is radiative only. On the other hand, the excellent conductivity of gold at THz frequencies prevents any transmission through a 100 nm gold layer (unlike in the IR spectral range), resulting in $\Gamma_{rad}^2 = 0$. Losses are then due to the gold mirror only and we can express:

$$Q_{DBR} = \frac{2\pi f_0}{\Gamma_{rad}^1} \quad \text{and} \quad Q_{gold} = \frac{2\pi f_0}{\Gamma_{diss}^2} \quad (3)$$

with $\Gamma_{rad}^1 = \frac{1}{\tau^*}(1 - |r_{DBR}|^2)$ and $\Gamma_{diss}^2 = \frac{1}{\tau^*}(1 - |r_{gold}|^2)$[27], $f_0$ being the cavity resonant frequency. $r_{DBR}$ is calculated using TMM[28] and $r_{gold}$ from Fresnel equations and the gold refractive index.[29] $\tau^*$ refers to the round-trip time inside the cavity, which is also the effective phase reflection delay described in section 3.B. Due to the importance of the round-trip time $\tau^*$, the use of $3\lambda_0/4n_{Si}$ thickness of the Si layers as a direct influence on the quality factor: without any influence on both mirror losses, the additional $\lambda_0/2n_{Si}$ Si thickness significantly increases the round-trip time inside the cavity and hence reduces the coupling and dissipation rates.[30] In the end, the quality factor for our THz Tamm cavity reads:

$$Q = 2\pi \frac{f_0}{(\Gamma_{rad}^1 + \Gamma_{diss}^2)} = 2\pi \frac{\tau^* f_0}{(1 - |r_{DBR}|^2) + (1 - |r_{gold}|^2)} \quad (4)$$

Total absorption is achieved when the radiative coupling and the dissipation rates are equals: it is the critical coupling criterion.[31] Thus, the minimum reflectivity is given by:

$$R_{min} = \left(\frac{\Gamma_{rad}^1 - \Gamma_{diss}^2}{\Gamma_{rad}^1 + \Gamma_{diss}^2}\right)^2 \quad (5)$$

We represent in Fig. 1(c), the reflectivity spectra computed using TMM of Tamm cavities for increasing silicon layers in the DBR. Adding silicon/vacuum pairs to the DBR increases $|r_{DBR}|$ and $Q_{DBR}$, resulting in the cavity quality factor $Q$ to converge to the ultimate limit of $Q_{gold}$ = 932. However, this leads to a dramatically reduced peak contrast due to deviation from critical coupling. To get more insight on this criterion, we represent in Fig. 1(d) the reflection coefficient of the cavity $r$, computed using TMM, in the complex plane.[32] Far from the resonance, the cavity just behaves as its DBR part (gray dotted line), having a reflection modulus $|r|$ close to 1 and a phase varying linearly inside the stop-band.[26] Hence, $r$ follows the unity circle, moving counterclockwise. Close to resonance, $|r|$ drops and $r$ describes a smaller inner circle (counter-clockwise), attached to the $r = -1$ point, of radius $\frac{2\Gamma_{rad}^1}{\Gamma_{rad}^1 + \Gamma_{diss}^2}$. Critical coupling is achieved when $r = 0$ is reached (dashed line circle). The 2 silicon layers cavity is then said to be overcoupled ($\Gamma_{rad}^1 > \Gamma_{diss}^2$) whereas the 3 and 4 silicon layers cavities are undercoupled ($\Gamma_{rad}^1 < \Gamma_{diss}^2$). Undercoupling is even larger for 5 and more layers cavities: each additional silicon/vacuum pair after the 4$^{th}$ silicon layer typically divides the resonance peak contrast by a factor ~10 due to the increasing DBR reflectivity. The 3 silicon layers case is almost critically coupled: this is the reason for the excellent contrast of the resonance represented Fig. 1(c) ($R_{min}$=3%). To meet the critical coupling, the tuning of $\Gamma_{rad}$ must be finer. Smaller DBR quality factor steps for additional dielectric pairs could be achieved by reducing the refractive index contrast between dielectric layers, using for instance polymers instead of vacuum.

With the optimal 3 silicon layers structure, we compute using TMM the electric field profile inside the cavity for the fundamental resonant mode at 1 THz (see Fig. 1(e)). The characteristic feature of the Tamm mode appears on this electric field profile: the amplitude maximum is located at the silicon-vacuum interface closest to the gold mirror. A second maximum observed inside the final silicon layer is only due to the $3\lambda_0/4n_{Si}$ thickness of the silicon layer and does not exist for a $\lambda_0/4n_{Si}$ thickness silicon layer. Owing to the high refractive index contrast between silicon and vacuum used in the THz DBR, the electromagnetic field is mainly confined in the first silicon-vacuum layers pair, leading to a $\lambda$ confinement length in the propagation direction, with $\lambda$ the effective wavelength in the medium. For a $\lambda_0/4n_{Si}$ thickness silicon layer, this electromagnetic field confinement is reduced down to $\lambda/2$ (as discussed in the **Supplementary S2**), which is the optimal value for propagation-based resonators.[33] The particular electric field distribution within THz Tamm cavities highlights their interesting properties. First, the coupling between the electric field and an active material is optimal for both 2D materials (interface 2-3 in Fig. 2(e)) and bulk materials (core of layer 2 in Fig. 2(e)). Graphene embedded within Tamm cavities using spacer layers has been previously theoretically studied in the IR range[34]. For Tamm cavity operating at THz frequencies, as the optimal 2D material position stands on a free silicon-air interface, the incorporation of a 2D material within the cavity is easily achievable. Indeed, it is compatible with direct transfer of usual fabrication techniques on silicon chip. Large coupling with bulk material is also an interesting feature, common to Fabry-Perot cavities but not provided by the majority of electronic resonators that relies only on the fringing fields that leak out from the circuit capacitor.[35] Second, the Si layer can act as a gate electrode for 2D active materials deposited at the silicon-vacuum interface (where electric field is maximum) to control their chemical potential without introducing any electromagnetic perturbation. Besides, the metallic layer is well-adapted for an electrical pumping of active bulk materials.

**Experimental Characterization**

We fabricate Tamm cavities as described above by stacking ~70 μm thick, double side polished, high resistivity silicon wafers (ρ>8000Ω.cm) separated by 75 μm spacers and a top gold mirror. Vacuum spacing (layers) are created by metallic strips of precise thickness intercalated between Si wafers. The top gold mirror is realized using thermal evaporation under vacuum of a 100 nm gold layer on the top silicon wafer. The assembly of the whole structure is realized manually using tweezers in a custom made sample holder, that includes grooves meant to maintain the metallic spacers and mechanical clamps ensuring the integrity of the structure. This overall simple procedure makes our THz Tamm cavity easy to fabricate. Since the first and final layers are necessarily Si layers, an N Si layers structure presents N-1 vacuum layers. A side view of a fabricated Tamm cavity made of 3 Si–2 vacuum layers is reported in Fig. 2(a). To characterize the electromagnetic modes in the THz Tamm cavity, we use a Fourier transform infrared (FTIR) spectrometer based on a Globar thermal light source and a liquid helium cooled bolometer detector. The spectral resolution of the FTIR spectrometer is 6 GHz.

The reflectivity spectra of a Tamm cavity with 3 Si layers (red line) and of the related DBR (black line - same structure without the gold mirror) are shown in Fig. 2(a). Whereas the DBR structure displays the usual stop-band, with a near unity reflectivity over a bandwidth of 0.3 THz and relatively low reflectivity outside the stop band, the reflection spectrum of the Tamm cavity exhibits a clear resonant mode at 0.915 THz. The resonant frequency lies in the center of the corresponding DBR stop-band, in agreement with TMM calculation. The quality factor deduced from these measurements is 121, limited by spectral resolution of the FTIR. Fig. 2(b) presents the measured reflectivity spectra of Tamm cavities with 2, 3 and 4 Si layers. As the number of Si layers increases from 2 to 3, we observe a narrowing of the resonant peak from 11 GHz to 7.5 GHz. Above 3 Si layers, the resonance linewidth becomes relatively constant because the resonance linewidth is too small to be resolved by the FTIR spectrometer. This limited spectral resolution of the measurement also reduces the contrast of the resonant peak for Tamm cavities with 3 and 4 Si layers. The slight shifts in the resonance frequency from one structure to the other is only due to small variations of the Si layers thickness. Higher order resonant Tamm modes can also be observed at 2.75 THz and 4.57 THz (for details see **Supplementary S3**).

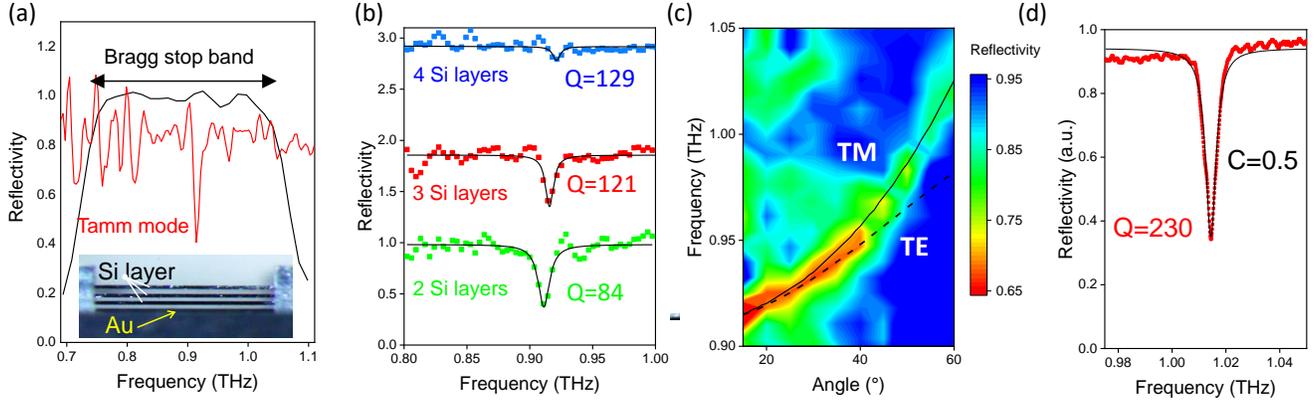

**Fig. 2.** (a) Experimental reflection spectrum of a 3 Si layers Tamm cavity, alongside reflection spectrum of a 3 silicon–2 vacuum layers DBR (FTIR measurements). The thickness of the Si layers for this cavity is 73 ± 2 μm. Inset: Side view photograph of a Tamm cavity, exhibiting vacuum gaps visible to the naked eye. (b) Experimental reflection spectra of 2 (green), 3 (red) and 4 (blue) Si layers Tamm cavities, with 6 GHz resolution (spectra are off-set by unity steps for clarity). The continuous black lines are Lorentzian fits giving quality factors of 84, 121 and 129 for 2, 3 and 4 Si layers Tamm cavities, respectively. (c) Dispersion relation versus incidence angle of the 3 Si layers cavity, from reflectivity measurements. Plain and dashed lines are the theoretical prediction of the dispersion relation for TM and TE polarization from TMM, respectively. (d) High resolution (0.6 GHz) reflection spectrum of a 3 Si layers Tamm cavity, measured using high resolution FTIR and synchrotron radiation. The thickness of the Si layers for this cavity is 66 ±2 μm. The FWHM of the Tamm resonance is 4.4 GHz corresponding to a quality factor of 230.

To characterize the in-plane dispersion properties of Tamm modes, we perform angle resolved reflectivity measurements from 15° to 60°. Fig. 2(c) shows the modal dispersion relation of a 3 Si layers Tamm cavity. We observe a parabolic dispersion, similarly to a Fabry-Perot mode. Being excited by free-space THz plane waves, the Tamm mode is radiative and lies within the light cone. TMM calculation reveals that the dispersion relation of Tamm modes formed by the TE and TM polarizations are distinct with a progressive frequency splitting between TE and TM polarizations as the in-plane wavevector is increased (dashed and solid lines, respectively, Fig. 2(c)). This frequency splitting is due to the polarization dependence of the reflections at the dielectric interfaces. Since the light emitted from the Globar is unpolarized, the experimental relation dispersion includes both dispersion relations of TM and TE polarized Tamm modes. However, due to the limited spectral resolution of the measurements, the frequency splitting between TM and TE polarization is not resolved at small angles. At angles larger than 40°, only the TM Tamm mode dispersion is observed in experimental data since the TE polarized mode peak becomes weaker and narrower. Besides, we verify that the Tamm dispersion relation well follows the predicted DBR relation dispersion, confirming that the DBR stop-band center dictates the resonance frequency of the Tamm cavity.

To experimentally determine the quality factor of a 3 Si layers Tamm cavity, we perform higher resolution reflectivity measurements using the SOLEIL synchrotron source. Measurements at the AILES beamline are performed on a Bruker IFS125 HR interferometer operating at 600 MHz resolution combined with synchrotron radiation. The reflectivity was measured at quasi normal incidence (7°) by means of a Helium pumped bolometer and a 50-μm thick beamsplitter.[36]

The THz reflection spectrum of a 3 Si layers cavity, reported in Fig. 2(d), shows a sharp resonance at 1.015 THz. The full-width at half-maximum (FWHM) of this Tamm cavity mode is ~4.4 GHz corresponding to a quality factor as high as 230. The contrast reaches 50 %. We attribute the lower experimental quality factor than predicted by ~x2 to inhomogeneous broadening of the Tamm mode due to local layer thickness variations (~100 nm) over the surface of the THz beam (~1 mm), as discussed in **Supplementary S4**. Further developments are in progress to overcome these fabrication issues. This high contrast value of the cavity resonance combined with a high-quality factor, makes the Tamm cavity especially suited to be coupled to low absorbing materials such as graphene. Indeed, by loading the cavity with a low absorbing material, the critical coupling and thus maximized coupling efficiency could be achieved. Besides, with Q>200, which lies within the same range of those reported in THz Fabry-Perot cavity[1,37], THz light-matter interaction is significantly enhanced in THz Tamm cavity, opening promising perspectives for the development of advanced THz optoelectronic devices such as sensitive detectors, high contrast modulators, narrow filters and lasers as well as for THz cavity quantum electrodynamics.

## TAMM CAVITY WITH SUB-WAVELENGTH METAL GRATING

### Experimental Results

An interesting feature of Tamm cavity is the possibility to pattern the metallic mirror providing a wide range of functionalities. Here, we focus on the patterning of the metallic mirror into a periodical gold subwavelength (sub-λ) grating and explore its potential for tuning the frequency of the THz Tamm mode.[38] The cavity is composed of a 3 high-resistivity Si wafers based DBR and a sub-λ periodic gold strip grating deposited on top of the DBR (Fig. 3(a)). The surface patterning of the gold mirror was realized using simple laser lithography directly on surface of the top Si wafer of the DBR, prior to the gold layer deposition, followed by a lift-off in acetone (see optical microscope image on Fig. 3(b)). For each sub-λ metal grating cavity, we preserve a plain gold area (without any patterning) to provide a reference Tamm cavity (fabricated with the same DBR). The relevant tuning parameter of this structure is the filling factor $ff$, i.e. the ratio between the strip width $a$ and the grating period $p$. Thus, $ff$ is the fraction of gold on the DBR surface and the reference Tamm cavity corresponds to $ff$=1. In the following, $p$ will be fixed to 75 μm ($\lambda_0/4$), $a$ being the parameter actually changing. This ensures the grating to remain subwavelength even in silicon, in which $\lambda=\lambda_0/n_{Si}$, preventing any grating mode resonances within the spectral range considered here[39,40]. The patterned mirror exhibits a non-negligible transmission, which leads to a reduction of the quality factor value (for more details see **Supplementary S5**). To tackle this issue, we add an additional full gold mirror at a distance $\lambda_0/4$ (75 µm) over the strip grating mirror, used to force the transmitted electromagnetic field back into the cavity. In the following, we will refer as "orthogonal" and "parallel" polarizations for electric field perpendicular and aligned to the strips, respectively.

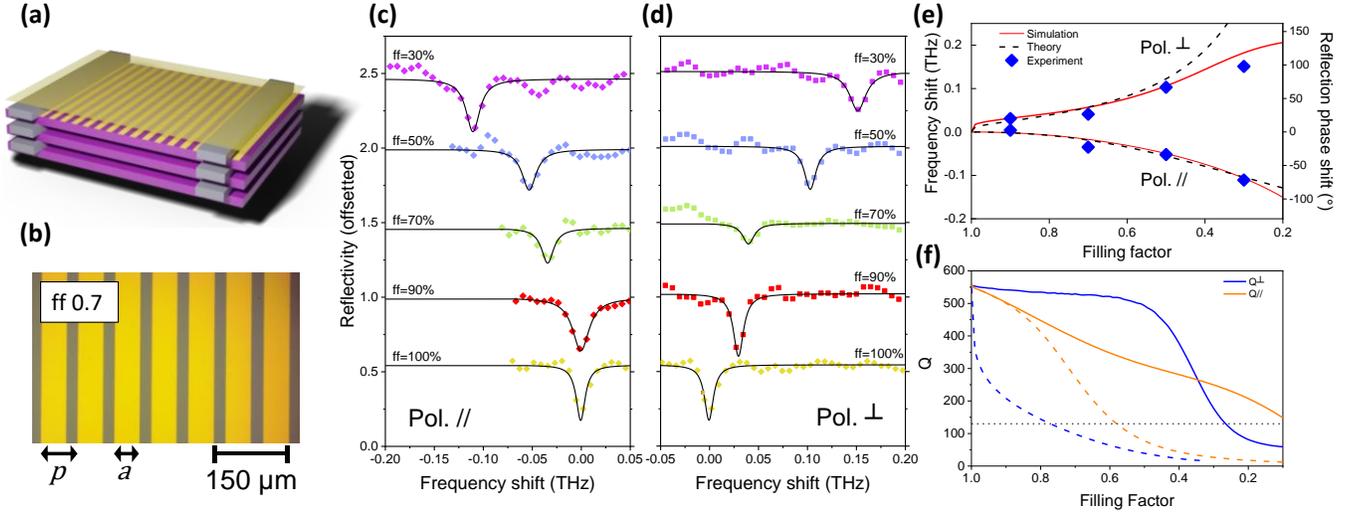

**Fig. 3.** (a) Schematic of the sub-wavelength metal grating Tamm cavity. Compared to Fig. 1(a), the gold mirror has been turned into a periodic strip grating, and an additional gold solid mirror stands at distance $\lambda_0/4$ (75µm) over the structure (transparent for visibility) (b) Optical microscope picture of the gold strip grating used as top mirror (ff=70%) (c) and (d) Experimental reflection spectra of Tamm cavities for decreasing filling factors, bottom to top, for a parallel and orthogonal polarization respectively. X-axis represents the frequency shift, $\Delta f$, with respect to the resonant frequency of the reference Tamm cavity (ff=1) placed on the same wafer. The curves are shifted with a constant offset for clarity. The experimental data is indicated by dotted curves and the continuous black lines are Lorentzian fits. (e) Frequency shifts of Tamm cavities from FEM simulations (red curve), theory (dashed black curve) and experiments (blue diamonds) as a function of the filling factor. Right scale indicates the corresponding reflection phase shift from theory (Section 2B). (f) Simulated quality factors (FEM) as a function of the filling factor for parallel (orange) and orthogonal (blue) polarizations. Solid lines: complete cavity with additional top mirror. Dashed lines: Cavity without top mirror. Dotted black line: maximum measurable quality factor corresponding to the resolution of our FTIR.

Fig. 3(c,d) show the measured reflectivity of subwavelength grating Tamm cavities (ff<1) for different filling factors ranging from 0.9 to 0.3 and for both parallel and orthogonal polarizations. The reflectivity is displayed as a function of the frequency shift from the resonant frequency of the reference Tamm cavity (ff=1). We observe a clear shift of the resonant frequency as the filling factor is decreased. The structure exhibits a total frequency tuning range >0.25 THz for ff=0.3, which is 50 times higher than the 4.4 GHz linewidth of the cavity mode presented in Fig. 2(d). We also observe that the resonant frequency shift, $\Delta f$, evolves in opposite directions depending on the polarization: for a parallel polarization, the resonant frequency shifts toward lower frequencies ($\Delta f<0$), whereas it shifts toward higher frequencies for the orthogonal polarization ($\Delta f>0$). Moreover, the frequency tuning is not symmetric, being more sensitive to the filling factor for the orthogonal polarization. This highlights the difference between physical mechanisms involved depending on the mode polarization. Since the maximum measurable Q is limited by the FTIR spectral resolution to <130 (dashed line Fig. 3(f)), we are unable to capture from these reflectivity measurements the quality factor dependence on the filling factor.

To quantitatively establish how the quality factor of the Tamm cavity evolves with the filling factor of the subwavelength metallic grating, we calculate the reflection spectra of the subwavelength grating Tamm cavities for various ff using FEM simulations (COMSOL Multiphysics). Figure 3e reports the frequency shift, $\Delta f$, as a function of ff, extracted from calculation (dashed blue lines) and experimental data (red diamonds) for parallel and orthogonal polarizations. The good agreement between simulated and experimental $\Delta f$ for each polarization validates our calculation. We then extract from the calculation the quality factor of the Tamm modes as a function of the filling factors of the subwavelength gratings as represented in Fig. 3(f) by the solid lines. We observe that the quality factor of 550 is remarkably almost constant for the orthogonal polarization from ff=1 down to ff=0.5. Thus, a high-quality factor is preserved for a large range of frequency tuning. The role of the top mirror to preserve high quality factor by compensating the high transmission of gold strip gratings for this polarization is evidenced by the blue dashed line in Fid 3.f. Indeed, the quality factor calculated without top mirror shows a large drop as soon as gaps are opened between strips (as soon as ff≠1). Below ff=0.5, the quality factor with top mirror for this orthogonal polarization drops rapidly due to the decreased reflectivity of the DBR at stop-band edges. In return, for the parallel polarization, the quality factor with top mirror is slowly reduced as ff is decreased but remains as high as 180 for ff of only 0.1. Similar evolution for filling factors close to 1 is observed without top mirror, but it drops relatively faster at ff<0.8. Simulation shows that the main dissipation process, in this parallel case and for ff>0.8, is ohmic losses from currents in the strips (See **Supplementary S6**), whereas grating transmission contribution to losses becomes dominant for lower ff. This important contribution of ohmic losses explains why the top mirror is not as efficient to maintain a high-quality factor for this parallel polarization.

Note that the maximum frequency tunability is limited by the spectral bandwidth of the DBR stop-band: since the DBR is not an efficient reflector outside of its bandwidth, there is no existence of a Tamm mode outside of it. The DBR bandwidth can be expressed as:[26]

$$\Delta f = f_0 \frac{4}{\pi} arcsin\left(\frac{n_H - n_L}{n_H + n_L}\right) \quad (6)$$

$f_0$ being the stop-band center frequency. Because of the high refractive index contrast between Si ($n_H = 3.42$) and vacuum ($n_L = 1$), the THz DBR stop-band is particularly large, higher than 0.7 THz for a $\lambda_0/4n_{Si}$ layers structure. Such high refractive index contrast is not possible for $\lambda_0/4$ $n_{Si}$ layers in the visible and near-IR range, making the THz Tamm cavity especially well-suited for this frequency tuning functionality. Note that as mentioned previously, in our case, the DBR stop-band is split in 3 sub-stop-bands due to the use of 3 $\lambda_0/4n_{Si}$ thick Si wafers. The center sub-stop-band, centered at 1 THz, is narrower, but still provides a full tunability over 0.35 THz.

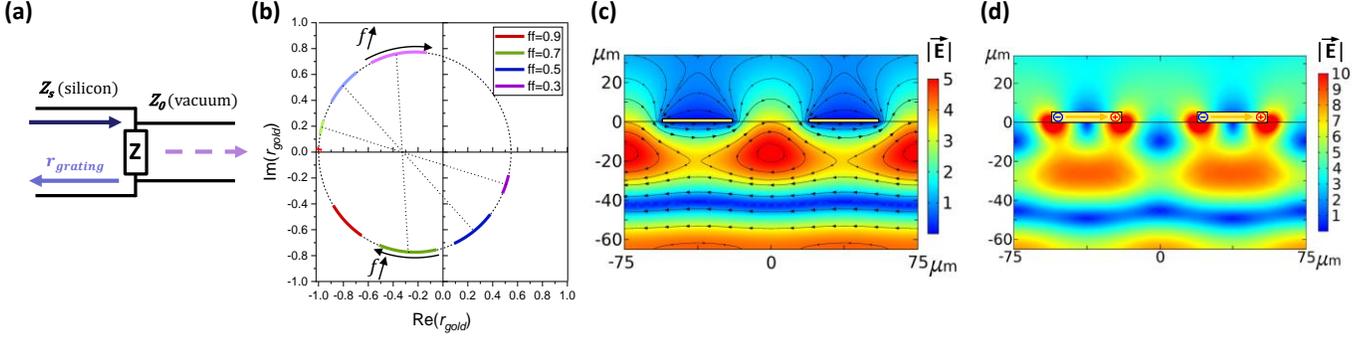

**Fig. 4.** (a) Schematic of the equivalent circuit model analogy. Vacuum is modeled as a transmission line of characteristic impedance $Z_0$, the impedance of free space (377 Ω), and silicon by a transmission line of impedance $Z_S = Z_0 \times n_{Si}$. An incoming wave inside silicon is partially reflected on the shunt impedance or strip grating and partially transmitted in vacuum. (b) Evolution of the reflection coefficient of the subwavelength strip grating in the complex plane versus filling factor and frequency (0.7 to 1.3 THz). Dotted diameters illustrate symmetry relations between grating of opposite polarizations and complementary filling factors. (c) and (d) Tamm mode electric field profiles for $ff$=0.5 strip grating cavities for parallel (f=0.941 THz) and orthogonal (f=1.102 THz) polarization, respectively. Field lines in (c) represents the local magnetic field direction and intensity. They are a signature of the magnetic field created by induced current in the strip in blue region around the strips.

**Theoretical Description**

To get insight into the different physical mechanisms involved in the frequency tuning for both polarizations, we develop a theoretical description of the Tamm cavity with subwavelength metal grating, based on an equivalent circuit model as described in Fig. 4(a). The resonance condition of a Tamm cavity is given by equation (1). The reflection phase of the DBR, $\arg(r_{DBR})$, is composed of the propagation contribution in the last Si layer (layer 2 in Fig. 1(e)), $2ke_{layer}$, and the reflection phase of the rest of the DBR (from 2-3 interface in Fig. 1(e)). Thus, inside the stop-band of center frequency $f_0$, $\arg(r_{DBR})$ can be approximated at first order by:[26]

$$\arg(r_{DBR}) = -2\pi\tau(f-f_0) + O((f-f_0)^3) - 2ke_{layer} \quad (7)$$

where $\tau$ is the phase reflection delay at the last vacuum-Si interface of the rest of the DBR (from 2-3 interface in Fig. 1(e)). For the last Si layer of $3\lambda_0/4n_{Si}$ thickness (layer 2 in Fig. 1(e)), $2ke_{layer} = 3\pi + 3\pi(f-f_0)T_0$, with $T_0 = 1/f_0$, being the DBR central period. Then:

$$\arg(r_{DBR}) = \pi - 2\pi(f-f_0)\left(\tau + \frac{3T_0}{2}\right) + O((f-f_0)^3) \quad (8)$$

This expression exhibits an effective phase reflection delay $\tau^* = \tau + \frac{3T_0}{2}$ for the whole DBR considered for our Tamm cavity, which is also the round trip-time inside the cavity. Using TMM, we calculate $\tau$=0.30 ps and $T_0$=1 ps. The phase of the reflection on the subwavelength periodic strip grating $\arg(r_{grating})$ can be expressed by modeling strip gratings as shunt impedance analytically derived from the actual strip grating geometry (Fig. 4(a)).[41,42,43] In this model, $r_{grating}$ is expressed as:

$$r_{grating} = \frac{Z(n_{Si}-1)-1}{1+Z(n_{Si}+1)} \quad (9)$$

where $Z$ is the reduced impedance (actual impedance normalized by vacuum impedance $Z_0$) of the strip grating. Depending on its polarization, the interaction of light with the metallic strip grating is completely different giving rise to distinct electric field distribution and two distinct values for $Z$. For a polarization parallel to the strips axis, the incident wave induces currents along the strips and creates magnetic field (black looped field lines around the strips, Fig. 4(c)), whereas a polarization orthogonal to the strips induces charge and hotspots at strip edges, resulting in electric field between the strips (Fig. 4(d)). These two pictures illustrate how the strip grating behaves as an inductance ($Z_{ind}$) or capacitor ($Z_{capa}$) for a parallel or orthogonal polarization. For a perfect metal, which is a reasonable approximation for gold in the THz range, these reduced impedances are given by:[42]

$$Z_{ind} = j\frac{p}{\lambda}\ln\left(\frac{1}{\sin\left(\pi\frac{ff}{2}\right)}\right) \text{ and } Z_{capa} = -j\frac{1}{2(1+n^2)}\frac{\lambda}{p}\frac{1}{\ln\left(\frac{1}{\cos\left(\pi\frac{ff}{2}\right)}\right)} \quad (10)$$

where $p$ is the strip grating period. These formulas only involve geometric parameters. The calculated real and imaginary part of the reflection coefficient of the strip grating for different filling factors are plotted on a Smith chart (see Fig. 4(b)). All values lie on a circle whose intersections with the real axis correspond to the limit case encountered: if $ff=1$, the incident wave is reflected on a perfect mirror (Z=0) and $r_{grating}=-1$; on the opposite, if $ff=0$, there is no gold on the silicon surface (Z=∞, there is no shunt) and the reflection coefficient is real, given by Fresnel equations: $r_{grating} = \frac{n_{Si}-1}{n_{Si}+1}$. Between these two points, increasing frequency lead to clockwise evolution along the circle for both polarizations. For perfect metals, Z is purely imaginary and the reflection phase can be reduced to a simple expression at first order:

$$\arg(r_{grating}) = \pi - 2nIm(Z) + O(Im(Z)^3) \quad (11)$$

The Taylor expansion is done for Im(Z) close to 0, i.e. for $ff$ close to 1. It is finally possible to derive using the phase matching criterion an analytical expression for the resonant frequencies of the strip grating Tamm cavity in the capacitive and inductive cases:

$$f_{ind} = f_0 \frac{1}{1+\frac{p}{c}\frac{\beta_{ind}(ff)}{2\pi\left(\tau+\frac{3T_0}{2}\right)}} \text{ and } f_{capa} = \frac{f_0}{2} + \frac{f_0}{2}\sqrt{1+\frac{2}{f_0^2}\frac{c}{p}\frac{\beta_{capa}(ff)}{2\pi\left(\tau+\frac{3T_0}{2}\right)}} \quad (12)$$

with:

$$\beta_{ind}(ff) = 2n\ln\left(\frac{1}{\sin\left(\frac{\pi ff}{2}\right)}\right) \text{ and } \beta_{capa} = \frac{n}{(1+n^2)}\frac{1}{\ln\left(\frac{1}{\cos\left(\pi\frac{ff}{2}\right)}\right)} \quad (13)$$

The validity of these formula is highlighted in Fig. 3(e) by the good agreement between the predicted resonant frequencies (dashed black curve) with the resonant frequencies obtained from FEM simulations (red curve) and the experimental results (blue points) down to $ff$~0.5. Small deviations at $ff$ close to 1 can be explained by the lossless nature of the analytical model, whereas FEM simulations takes into account actual complex refractive index of gold from.[299] This analysis is powerful for the design of subwavelength grating Tamm cavity by predicting with simple analytical expression the resonant frequencies for both polarizations.

From this model, we emphasize that the Tamm cavity resonant frequency shift is a direct signature of the reflection phase shift $\Delta\varphi = \arg(r_{grating}) - \pi$ due to the subwavelength structuration of the metallic layer given by:

$$\Delta\varphi = 2\pi(f - f_0)\left(\tau + \frac{3T_0}{2}\right) + O((f - f_0)^3) \quad (14)$$

$\Delta\varphi$ is then proportional to the Tamm cavity frequency shift $(f - f_0)$, as long as the resonant frequency does not reach the DBR stop band edges. From our experimental measurements, we can thus estimate the corresponding experimental reflection phase shifts $\Delta\varphi$ (Right scale in Fig. 3(e)). The cavity can be seen as a local phase probe, preventing the necessity of complex delay or interferometric measurements. This method could easily be extended to other metamaterial systems that can be processed on thin Si wafers.

## Conclusion

In conclusion, we report on a new Tamm cavity in the THz spectral range made of a distributed Bragg reflector and a metallic mirror, with a high-quality factor of 230 at ~1THz. Electromagnetic simulations reveal a mode confinement of λ/2 within the DBR and an ideal coupling of the Tamm modes to both bulk and 2D materials. By a subwavelength structuration of the metal layer, a fine-tuning of the Tamm mode over >250 GHz is achievable while preserving a relatively high-quality factor. Moreover, the Tamm mode becomes polarization sensitive. These THz Tamm cavities are very promising as basic building blocks of lasers, as basic components for the development of advanced THz devices such as sensitive detectors, high-contrast modulators, polarization-sensitive absorbers, narrow filters and for THz cavity quantum electrodynamics in nanostructures such as those based on 2D materials.

**Supporting information.** The Supporting Information is available free of charge via the internet at http://pubs.acs.org:

Influence of $3\lambda_0/4n_{Si}$ silicon layer thickness on reflectance spectra and electric field profile in Tamm cavities, Higher order modes experimental measurements, Modeling of mode broadening due to thickness variations, Influence the additional top mirror on strip grating Tamm cavity mode, Loss sources in strip grating Tamm cavity.

**Funding.** This project has received funding from the European Research Council (ERC) under the European Union's Horizon 2020 research and innovation program (Grant Agreement 820133).

**Acknowledgment.** We thank Aloyse Degiron, J.F. Lampin and R. Colombelli for valuable discussions and Jose Palomo for his help with the fabrication.

**Disclosure.** The authors declare no competing financial interest.

## REFERENCES

1. Q. Zhang, M. Lou, X. Li, J. L. Reno, W. Pan, J. D. Watson, M. J. Manfra & J. Kono, "Collective non-perturbative coupling of 2D electrons with high-quality-factor terahertz cavity photons" Nature Phys. **12,** 1005–1011 (2016).

2. G. Scalari, C. Maissen, D. Turčinková, D. Hagenmüller, S. De Liberato, C. Ciuti, C. Reichl, D. Schuh, W. Wegscheider, M. Beck, and J. Faist, "Ultrastrong Coupling of the Cyclotron Transition of a 2D Electron Gas to a THz Metamaterial", Science **335**, 1323 (2012).

3. C. G. Derntl, D. Bachmann, K. Unterrainer, and J. Darmo, "Disk patch resonators for cavity quantum electrodynamics at the terahertz frequency", Opt. Express **25**, 12311-12324 (2017).

4. Y. Todorov, A. M. Andrews, I. Sagnes, R. Colombelli, P. Klang, G. Strasser, and C. Sirtori, "Strong Light-Matter Coupling in Subwavelength Metal-Dielectric Microcavities at Terahertz Frequencies", Phys. Rev. Lett. **102**, 186402 (2009).

5. B. Paulillo, S. Pirotta, H. Nong, P. Crozat, S. Guilet, G. Xu, S. Dhillon L. H. Li A. G. Davies, E. H. Linfield and R. Colombelli, "Ultrafast terahertz detectors based on three-dimensional meta-atoms" Optica, **4**, 1451 (2017)

6. M. Jeannin, G. Mariotti Nesurini, S. Suffit, D. Gacemi, A. Vasanelli, L. H. Li, A. G. Davies, E. H. Linfield, C. Sirtori, and Y. Todorov, "", ACS Photonics **6**, 1207–1215 (2019).

7. D. Palaferri, Y. Todorov, Y. N. Chen, J. Madeo, A. Vasanelli, L. H. Li, A. G. Davies, E. H. Linfield, and C. Sirtori, "Patch antenna terahertz photodetectors" Appl. Phys. Lett. **106**, 161102 (2015).

8. K. Okamoto, K. Tsuruda, S. Diebold, S. Hisatake, M. Fujita, and Tadao Nagatsuma, "Terahertz Sensor Using Photonic Crystal Cavity and Resonant Tunneling Diodes", J. Infrared Milli. Terahz Waves **38**, 1085–1097 (2017).

9. G. Liang, X. Hu, X. Yu, Y. Shen, L. H. Li, A. Giles D., E. H. Linfield, H. K. Liang, Y. Zhang, S. F. Yu, and Q. J. Wang, "Integrated Terahertz Graphene Modulator with 100% Modulation Depth", ACS Photonics **2**, 1559 (2015).

10. B. Vasić and R. Gajić, "Tunable Fabry–Perot resonators with embedded graphene from terahertz to near-infrared frequencies", Opt. Lett. **39**, 6253-6256 (2014).

11. X. Wang, X. Jiang, Q. You, J. Guo, X. Dai, and Y. Xiang, "Tunable and multichannel terahertz perfect absorber due to Tamm surface plasmons with graphene," Photon. Res. **5**, 536-542 (2017).

12. R. Jago, T. Winzer, A. Knorr, and E. Malic, "Graphene as gain medium for broadband lasers", Phys. Rev. B **92**, 085407 (2015).

13. D. Yadav, G. Tamamushi, T. Watanabe, J. Mitsushio, Y. Tobah, K. Sugawara, A. A. Dubinov, A. Satou, M. Ryzhii, V. Ryzhii and T. Otsuji, "Terahertz light-emitting graphene-channel transistor toward single-mode lasing", Nanophotonics **7**, 741–752 (2018).

14. F. Hindle, R. Bocquet, A. Pienkina, A. Cuisset, and G. Mouret, "Terahertz gas phase spectroscopy using a high-finesse Fabry–Pérot cavity," Optica **6**, 1449-1454 (2019).

15. D.W. Vogt, A.H. Jones, R. Leonhardt, "Terahertz gas-phase spectroscopy using a sub-wavelength thick ultrahigh-Q microresonator, " Sensors, **20**, 3005 (2020).

16. Q. Lu, X. Chen, C.-L. Zou, and S. Xie "Extreme terahertz electric-field enhancement in high-Q photonic crystal slab cavity with nanoholes", Opt. Express **26** 30851-30861 (2018).

17. G. Lheureux, S. Azzini, C. Symonds, P. Senellart, A. Lemaître, C. Sauvan, J-P Hugonin, J-J. Greffet, and J. Bellessa, "Polarization-Controlled Confined Tamm Plasmon Lasers", ACS Photonics **2**, 842–848 (2015).

18. A. Kavokin, I. Shelykh, and G. Malpuech "Optical Tamm states for the fabrication of polariton lasers", Appl. Phys. Lett. **87**, 261105 (2005).

19. M. E. Sasin, R. P. Seisyan, M. A. Kalitteevski, S. Brand, R. A. Abram, J. M. Chamberlain, A. Yu. Egorov, A. P. Vasil'ev, V. S. Mikhrin, and A. V. Kavokin, "Tamm plasmon polaritons: Slow and spatially compact light", Appl. Phys. Lett. **92**, 251112 (2008)

20. M. Lopez-Garcia, Y.-L. D. Ho, M. P. C. Taverne, L.-F. Chen, M. M. Murshidy, A. P. Edwards, M. Y. Serry, A. M., Adawi, J. G. Rarity, and R. Oulton, "Efficient out-coupling and beaming of Tamm optical states via surface plasmon polariton excitation", Appl. Phys. Lett. **104**, 231116 (2014).

21. C. Symonds, S. Azzini , G. Lheureux, A. Piednoir, J. M. Benoit, A. Lemaitre , P.Senellart, and J. Bellessa, "High quality factor confined Tamm modes", Scientific Reports **7**, 3859 (2017).

22. S. Azzini, G. Lheureux, C. Symonds, J.-M. Benoit, P. Senellart, A. Lemaitre, J.-J. Greffet, C. Blanchard, C. Sauvan, and J. Bellessa, "Generation and Spatial Control of Hybrid Tamm Plasmon/Surface Plasmon Modes", ACS Photonics 2016, **3**, 1776–1781 (2016).

23. J. M. S. S. Silva and M. I. Vasilevskiy, "Far-infrared Tamm polaritons in a microcavity with incorporated graphene sheet", Optical Materials Express **9**, 244-255 (2019).

24. T. W. Du Bosq, A. V. Muravjov and R. E. Peale, "High-reflectivity intracavity Bragg mirrors for the far-infrared p-Ge laser", Terahertz for Military and Security Applications **2**, 167-173 (2004).

25. M. Kaliteevski, I. Iorsh, S. Brand, R. A. Abram, J. M. Chamberlain, A. V. Kavokin, and I. A. Shelykh, "Tamm plasmon-polaritons: Possible electromagnetic states at the interface of a metal and a dielectric Bragg mirror", Phys. Rev. B **76**, 165415 (2007).

26. D. I. Babic and S. W. Corzine, "Analytic expressions for the reflection delay, penetration depth, and absorbance of quarter-wave dielectric mirrors, "Journal of Quantum Electronics **28**, 514-524 (1992).

27. Z.-Y. Yang, S. Ishii, T. Yokoyama, T. D. Dao, M.-G. Sun, P. S. Pankin, I. V. Timofeev, T. Nagao, and K.-P. Chen, "Narrowband Wavelength Selective Thermal Emitters by Confined Tamm Plasmon Polaritons", ACS Photonics, 4, 9, 2212–2219 (2017) .

28. P. Yeh, A. Yariv, and C.-S. Hong, "Electromagnetic propagation in periodic stratified media. I. General theory*," J. Opt. Soc. Am. **67**, 423-438 (1977).

29. M. A. Ordal, Robert J. Bell, R. W. Alexander, L. L. Long, and M. R. Querry, "Optical properties of fourteen metals in the infrared and far infrared: Al, Co, Cu, Au, Fe, Pb, Mo, Ni, Pd, Pt, Ag, Ti, V, and W.," Appl. Opt. **24**, 4493-4499 (1985).

30. L. Siegman, "Lasers", University science books, Mill Valley, 428-430 (1986).

31. B. Auguié, A. Bruchhausen and A. Fainstein, "Critical coupling to Tamm plasmons", Journ. of Opt. **17**, 035003 (2015).

32. H. H. Haus, "Waves and fields in optoelectronics", Prentice-Hall, 204-206 (1984).

33. R. Coccioli, M. Boroditsky, K. W. Kim, Y. Rahmat-Samii, and E. Yablonovitch, "Smallest possible electromagnetic mode volume in a dielectric cavity." IEE Proceedings-Optoelectronics **145,** 391-397 (1998).

34. Hua Lu, Xuetao Gan, Baohua Jia, Dong Mao, and Jianlin Zhao, "Tunable high-efficiency light absorption of monolayer graphene via Tamm plasmon polaritons", Opt. Letters **41**, 20, 4743-4746 (2016).

35. D. Dietze, A. Benz, G. Strasser, K. Unterrainer, and J. Darmo, "Terahertz meta-atoms coupled to a quantum well intersubband transition," Opt. Express **19**, 13700–13706 (2011).

36. Roy, P., Rouzieres, M., Qi, Z.M., Chubar, O. "The AILES Infrared Beamline on the third generation Synchrotron Radiation Facility SOLEIL" Infrared Physcics and Technology., **49**, 139-146 (2006).

37. F. Meng, M. D. Thompson, B. Klug, D Čibiraitè, Q. Ul-Islam, and H. G. Roskos, "Nonlocal collective ultrastrong interaction of plasmonic metamaterials and photons in a terahertz photonic crystal cavity", Opt. Express **27**, 24455 (2019).

38. A. R. Gubaydullin, C. Symonds, J.-M. Benoit, L. Ferrier, T. Benyattou, C. Jamois, A. Lemaître, P. Senellart, M. A. Kaliteevski, and J. Bellessa, "Tamm plasmon sub-wavelength structuration for loss reduction and resonance tuning", Appl. Phys. Lett. **111**, 261103 (2017).

39. V. O. Byelobrov, T. L. Zinenko, K. Kobayashi and A. I. Nosich, "Periodicity Matters: Grating or lattice resonances in the scattering by sparse arrays of subwavelength strips and wires.",IEEE Antennas and Propagation Magazine **57**, 6, 34-45, (2015)

40. L. Baldassarre, M. Ortolani, A. Nucara, P. Maselli, A. Di Gaspare, V. Giliberti, and P. Calvani, "Intrinsic linewidth of the plasmonic resonance in a micrometric metal mesh", Opt. Express **21**, 15401-15408 (2013).

41. R. Ulrich, T. J. Bridges, and M. A. Pollack, "Variable Metal Mesh Coupler for Far Infrared Lasers," Appl. Opt. **9**, 2511-2516 (1970).

42. L. B. Whitbourn and R. C. Compton, "Equivalent-circuit formulas for metal grid reflectors at a dielectric boundary," Appl. Opt. **24**, 217-220 (1985).

43. O. Luukkonen, C. Simovski, G. Granet, G. Goussetis, D. Lioubtchenko, A. V. Räisänen, and S. A. Tretyakov, "Simple and Accurate Analytical Model of Planar Grids and High-Impedance Surfaces Comprising Metal Strips or Patches", IEEE Transactions on Antennas and Propagation **56,** 1624-1632 (2008).